\documentclass[%
reprint,
superscriptaddress,
nofootinbib,
nobibnotes,
 amsmath,amssymb,
aps,
prc,
]{revtex4-2}
\usepackage{amssymb}
\usepackage{graphicx}
\usepackage{dcolumn}
\usepackage{bm}
\usepackage{hyperref}
\usepackage{breakurl}

\begin{document}

\preprint{Preprint No. 1}

\title{On the randomness and correlation in the trajectories of alpha particle emitted from ${}^{241}$Am: Statistical inference based on information entropy}

\author{M. {El~Ghazaly}}
 \address{Department of Physics, Faculty of Science, Zagazig University, Zagazig, 44519, Egypt.}
\author{Elsayed K. Elmaghraby}
 \address{Experimental Nuclear Physics Department, Nuclear Research Centre, Egyptian Atomic Energy Authority, Cairo 13759, Egypt. Corresponding author.}
\email[Institutional email: ]{elsayed.elmaghraby@eaea.org.eg}
\email[Alt. email: ]{e.m.k.elmaghraby@gmail.com}
\author{A. Al-Sayed}
\address{Department of Physics, College of Science and Arts, Al-methnab, Qassim University, P. O. Box 931, Buridah 51931, Al-mithnab, KSA} \address{Department of Physics, Faculty of Science, Zagazig University, Zagazig, 44519, Egypt.}
\author{Amal Mohamed}
 \address{Department of Physics, Faculty of Science, Umm Al-Qura University, Makkah, Saudi Arabia.}
 \address{Department of Physics, Faculty of Science, Zagazig University, Zagazig, 44519, Egypt.}
\author{Mahmoud S. Dawood}
 \address{Department of Physics, Faculty of Science, Zagazig University, Zagazig, 44519, Egypt.}

\date{\today}

\begin{abstract}
Most particle detectors are based on the hypothesis that particles are emitted randomly upon nuclear decay. In the present work, we tested the hypothesis of the existence of correlation in the random trajectories of alpha particles emitted from ${}^{241}$Am source and the null hypothesis of random trajectories. The trajectories were clued through the registration of track in a solid-state nuclear track detector. The experimental parameters were optimized to identify the possible sources of correlation in the track registration and the detector conditions upon exposure and etching to avoid misleading results. The optimization included authentication of linearity in registration efficiency with exposure time to prevent coalescence of registered tracks. The statistical inference processes were based upon adaptive quadrates analysis of the spatial data, and entropy and divergence analysis of the quadrate data together with the null hypothesis of Poisson distribution of random trajectories. The clustering and dispersion analysis were performed with central deviation tendency, empirical K-function, radial distribution analysis, and proximity Analysis. Results showed a pattern of gained information within the registered tracks that may be attributed to {the alteration in the alpha particles' trajectories induced by the strong electric field due to atoms in the source compound and encapsulation film}.
\end{abstract}

\maketitle


\section{Introduction}
A few decades ago, numerous passive detectors, such as solid-state nuclear track detectors (SSNTD), were utilized to detect heavy ions from different radiation sources, including cosmic rays, accelerators, and laser-matter interactions \cite{Durrani1987book}. The most common SSNTD is poly allyl diglycol carbonate (PADC), commonly known as Colombia Rains PADC. PADC detector is characterized by a high registration efficiency that reaches 100\% for perpendicularly incident alpha particles. PADC tracks range from several micrometers to several nanometers for scanning with a high spatial resolution imaging system such as atomic force and confocal microscopes. Meanwhile, spatial resolution amounts to several micrometers for scanning with an ordinary optical microscope \cite{Durrani1987book,Galbraith2005Statistics,Nikezic200451,Azooz20122470,Elghazaly2018432,Nikezic20081417}. The major disadvantage of SSNTD is the time resolution of the PADC detector; it is impossible to discern the occurrence of the tracks in the PADC detector. Because SSNTD is a position-sensitive tool with a small spatial resolution, if the etching conditions were adjusted correctly, it can be used for the pattern of the particle beam. The basic idea of SSNTD is the damage zone along the trajectory of the incident particle through the detector's material (known as latent tracks) which are of some nanometers. The latent tracks in the PADC detector can be enlarged to be visible under an ordinary optical microscope by chemical etching at a proper concentration and temperature in NaOH or KOH aqueous solution\cite{Nikezic20081417}.

Alpha particles from ${}^{241}$Am, have enough energy to overcome the air deceleration and reach the surface of the SSNTD, inducing the latent track. The alpha-decay energies are associated with $^{241} $Am are  {5485.56} KeV for 84.8\% of the emitted particle, 5442.8 KeV for 13.1\% , and 5388 KeV for the remaining 1.6\% \cite{Basunia20062323}. {The isotropic nature of the alpha decay from $^{241} $Am is based upon the fact that one could not ``stimulate'' atomic nuclei to decay in specific direction; i.e. there is no preferred direction of the emitted alpha particles}. It could be determined statistically by the number of nuclei among a large assembly of them that would decay in a given time interval. Accordingly, alpha particles should be distributed randomly in a given area if the null hypothesis is correct; furthermore, it should follow complete spatial randomness (CSR).

Previous studies (cf. Refs. \cite{Tommasino1997Inbook,Yu200493,Hamza2008343,Lee2018192,Sohrabi201769,Yousef201723,Cockcroft2009conf,Zaki2017272,Zaki2007567}) had shown that the distribution of alpha particle tracks was neither uniform nor random, whatever the precaution made to make such a fact obvious. The reason may be any process from the decay to the registration on the detector surface. Accordingly, the present work aims to study the spatial distribution and correlation between alpha particle tracks from the ${}^{241}$Am source.

\section{Material and Method}
PADC detectors (Tastrak PADC; density = 1.32 g/cm${}^{3}$, molecular composition C${}_{12}$H${}_{18}$O${}_{7}$) of a thickness (500$\pm$4) $\mu$m were exposed to vertically incident alpha particle from $^{241} $Am (main alpha particle energy 5.486 MeV and having activity 9 $\mu $Ci in air). The vertical alignment was essential to eliminate, even, the small effect of gravity. Predefined average energy could be obtained between 0.5 and 5.5 MeV by changing the length of the air column between the $^{241} $Am source and the PADC detector with the aid of a collimator. Air molecules are moving randomly at room temperature so that the scattering shall be random. The $^{241} $Am source is coated by a 30 nm-thick layer of gold to avoid source corrosion and neutralize the effect of the recoil of $^{237}$Np atom and the electron negative charge\footnote{The alpha decay changes the energy states of the atom so that the $^{237}$Np atom reconfigures the binding levels for the 95-2 electrons. This charge reconfiguration process is complex process \cite{Lumpkin2012563}.  The two extra electrons will be left free in the lattice, then that bulk will be negatively charged.} remaining in the source material upon alpha particle emission \cite{Wiss2012465}. {The length of the collimator was 1.5 cm which is suitable to reduce the average of 5.5 MeV alpha particles to about 4 MeV of energy}. The cross-section diameter of the collimator was 1.5 mm (area of 1.77 mm${}^{2}$) to ensure that {most of} the alpha particle is perpendicularly striking the detector surface; such a narrower collimator minimizes the energy's spreading out due to different paths in air. The alpha-irradiated PADC detectors were etched in an aqueous solution of 6.25N NaOH at 70$\pm$1${}^{\circ}$C for different durations. The bulk-etching rate (\textit{V}${}_{B}$) was measured using the well-known weight and thickness decrement methods. It amounts to (1.26$\pm$0.06) $\mu $m/h. The track density and diameters were measured with an optical microscope (Nikon, ECLIPSE, E200) equipped with Nikon digital camera (DS-Fi1). All photomicrographs are captured in RBG colors with a dynamic range of 2${}^{16}$. RBG photomicrographs were analyzed using ImageJ software \cite{SchneiderIMAGEJ}, where different color channels are separated, and Green-channel is selected since it has maximum contrast.

\begin{figure*}
 \centering
\includegraphics[width=0.90\linewidth]{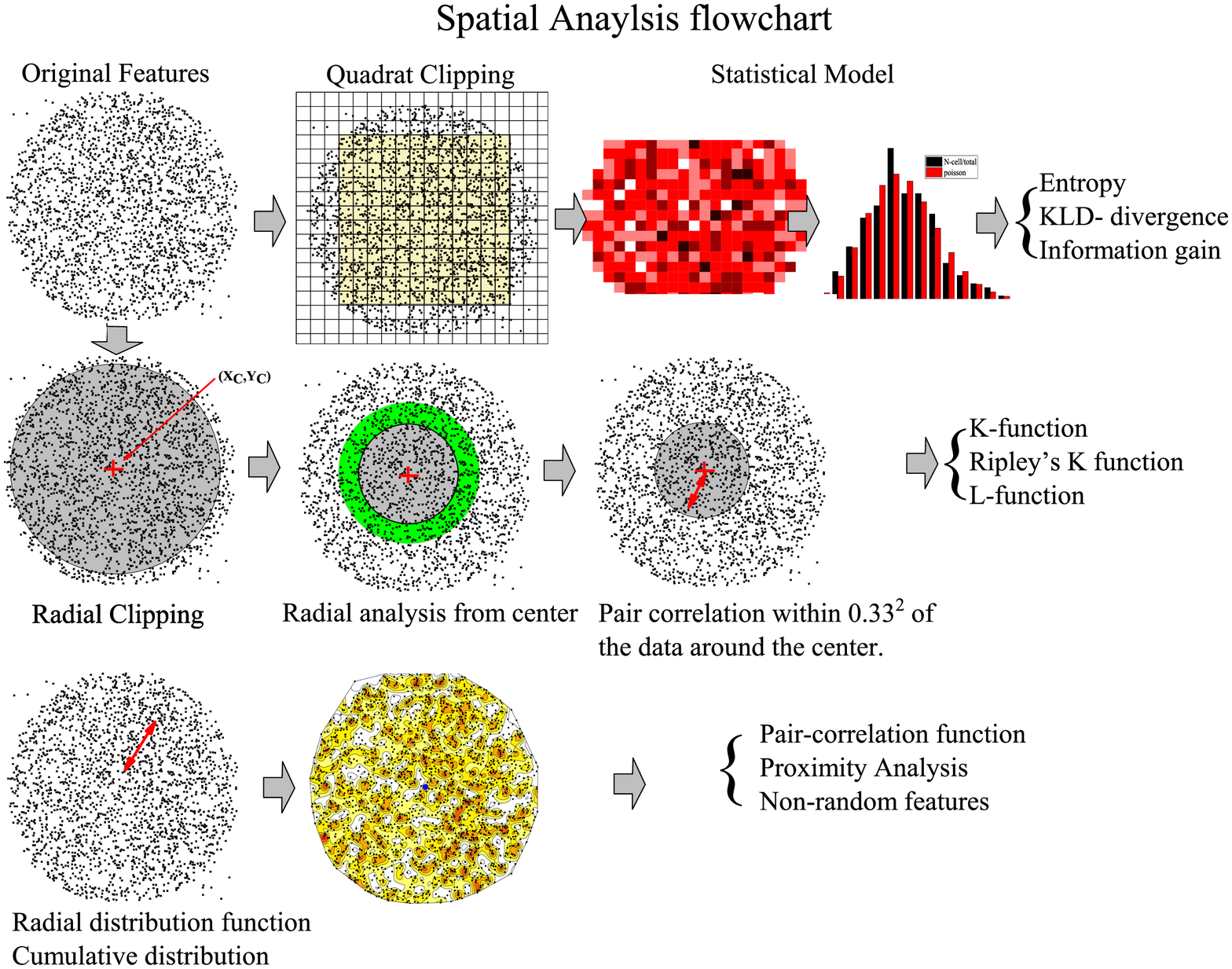}
 \caption{Flowchart of the spatial data analysis.}\label{fig5}
\end{figure*}

We apply three data reduction techniques and analysis on the track pattern registered on the PADC detector: (1) Divergence analysis between the null hypothesis of randomness and a test hypothesis of the registered pattern. (2) Dispersion analysis involving the central tendency and (3) Proximity analysis. {Fig}. \ref{fig5} shows the abstraction of the techniques used.

\section{Results and discussion}

\subsection{Verification of linearity in registration efficiency. }

The registration efficiency depends on the energy of the alpha particles reaching the surface of the detector. {For the source-to-detector distance of 1.5 cm}, the alpha particle reaches the surface with an average energy of 4 MeV. Several factors must be optimized to adjust the condition for linearity in registration efficiency: (1) The range of alpha particles having an energy of 4 MeV in the PADC detector is about 20.5 $\mu$m \cite{Ziegler198593book,Ziegler20101818}. The etching time ($t_{e} $) should be much less than the etching time to reach track depth at 20.5 $\mu$m (denoted $t_{R} $), which leads to detecting all incident alpha particles on the PADC detector \cite{Zylstra201284}. The minimal etching time will cause a minor track diameter and lower the detectability of the track using an ordinary optical microscope. (2) Another requirement is to adjust the registration efficiency associated with the coalescence of registered tracks when the fluence exceeded a specific limit, i.e., determination of the maximum number of tracks registered per unit area to maintain linearity between track density and exposure time. (3) Normality test for the registered tracks upon optimized conditions.

\begin{figure}
 \centering
\includegraphics[width=\linewidth]{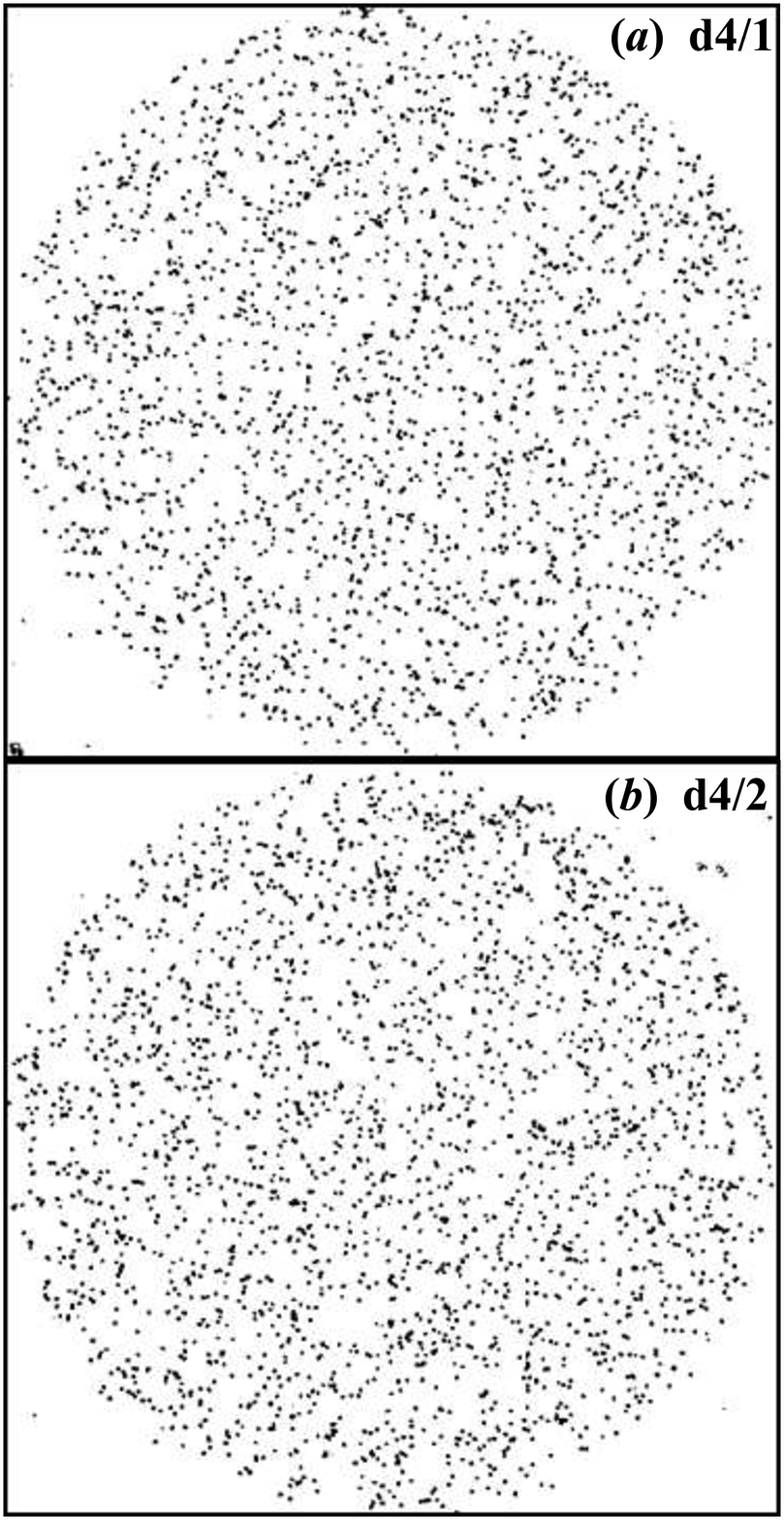}
\caption{Comparison between the distribution of alpha particles tracks free areas in two photomicrographs (a) for sample b4/1 and (b) for sample b4/2 of the PADC detector irradiated with 4 MeV alpha particle for 1 minutes with a time interval of ten minutes, samples were chemically etched in 6.25N NaOH at 70 $^\circ$C for 4 hours. The area of each picture is 1583$\times$1583 $\mu$m${}^{2}$.}\label{Fig1}
\end{figure}

Three sets of samples were exposed to alpha particles for durations between 30 s and 300 s; the first set was etched chemically in 6.25N NaOH at 70 $^\circ$C for 2 h. The second and the third set were etched in the same chemical conditions but for 4 h, and 6 h, respectively, see Table \ref{Tab1}. Images were recorded through the collimated area up to 1583$\times$1583 $\mu$m${}^{2}$. Since there was a need to optimize the detectible track diameter, the track size distribution was determined from the two pre-samples after being chemically etched in 6.25N NaOH at 70 $^\circ$C for 4 hours, see Fig. \ref{Fig1}. Despite both photomicrographs being produced under the same conditions, namely exposure time, etching time, alpha particle energy, and PADC detector, one can recognize the randomness in the registered patterns and the extension of the registration area to cover all the 1583$\times$1583 $\mu$m${}^{2}$ area. This assures that the alpha particle source is isotropically distributed behind this area.

\begin{table}[htb]
 \centering
 \caption{List of investigated samples and exposure and etching conditions. Letters will be used for the figure's label. $\hat{\lambda }$ is the registered track density in cm$^{-2} $. All samples were etched in 6.25N NaOH at 70 $^\circ$C.}\label{Tab1}
\begin{tabular}{l|c|c|c|c|c|c} \hline
 & \multicolumn{6}{c}{Etch time} \\ \cline{2-7}
Exposure & \multicolumn{2}{c}{2 h} & \multicolumn{2}{|c|}{4 h} & \multicolumn{2}{c}{6 h} \\ \cline{2-7}
time & Label & $\hat{\lambda }$ & Label & $\hat{\lambda }$ & Label & $\hat{\lambda }$ \\ \cline{1-7}
30 s & a2 & 7.58$\times 10^{4} $ & a4 & 7.31$\times 10^{4} $ & a6 & 6.03$\times 10^{4} $ \\
60 s & b2 & 1.39$\times 10^{5} $ & b4/1 & 1.21$\times 10^{5} $ & b6 & 1.07$\times 10^{5} $ \\
60 s & - & & b4/2 & 1.31$\times 10^{5} $ & - & \\
60 s & - & & b4/3 & 1.17$\times 10^{5} $ & - & \\
120 s & c2 & 2.58$\times 10^{5} $ & c4 & 2.61$\times 10^{5} $ & c6 & 1.44$\times 10^{5} $ \\
180 s & d2 & 3.68$\times 10^{5} $ & d4/1 & 2.81$\times 10^{5} $ & d6 & 1.27$\times 10^{5} $ \\
180 s & & & d4/2 & 2.82$\times 10^{5} $ & & \\
240 s & e2 & 4.89$\times 10^{5} $ & e4 & 4.27$\times 10^{5} $ & e6 & 1.39$\times 10^{5} $ \\
300 s & f2 & 5.16$\times 10^{5} $ & f4 & 5.02$\times 10^{5} $ & f6 & 1.10$\times 10^{5} $ \\ \hline\hline
\end{tabular}
\end{table}

\begin{figure}
 \centering
\includegraphics[width=\linewidth]{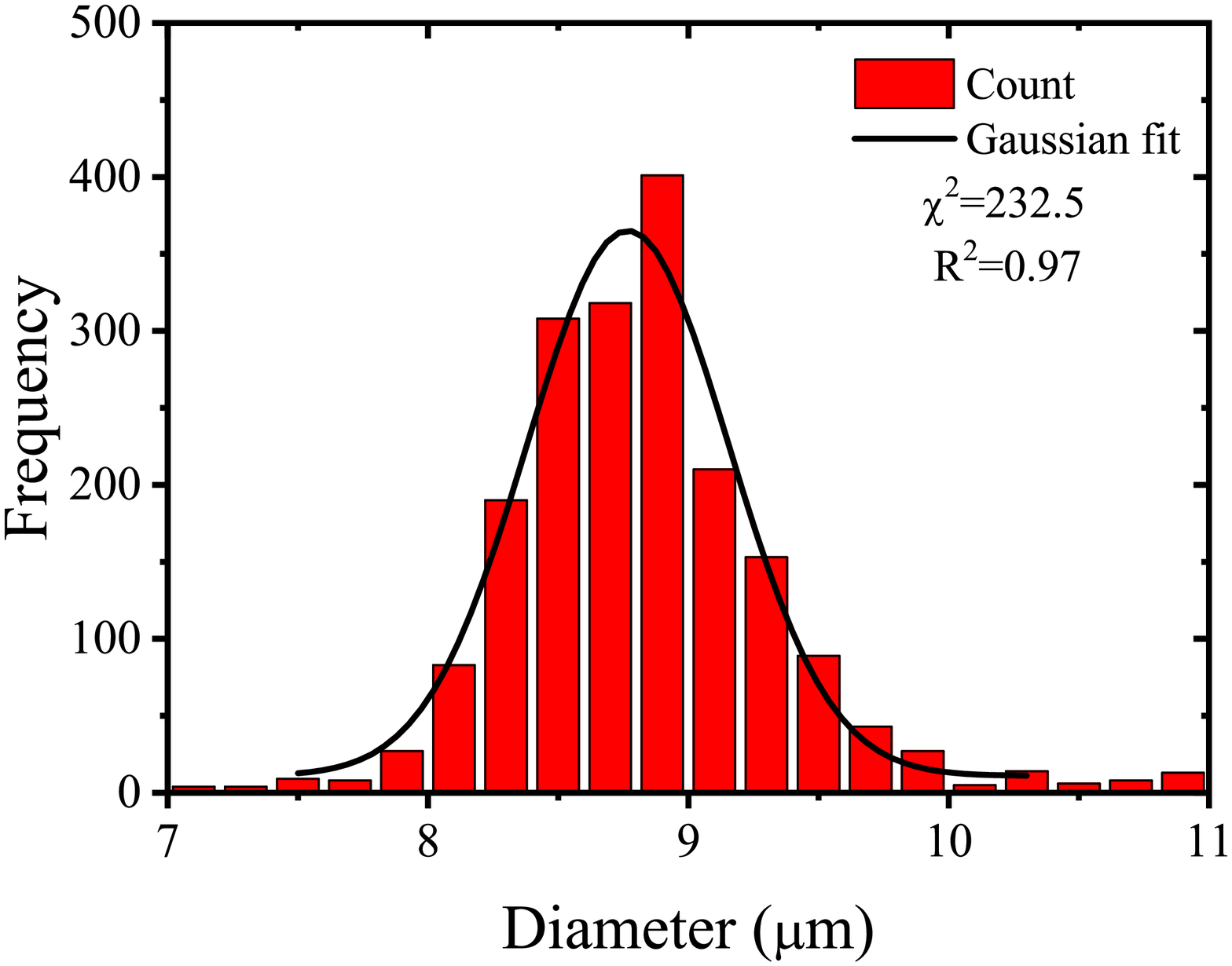}
 \caption{Histogram of the alpha tracks diameter in the PADC detector for samples d4/1 and d4/2.}\label{fig2}
\end{figure}

The track diameter distribution histogram in Fig. \ref{fig2} shows that the alpha particle track diameter follows Gaussian distribution centered at (8.77$\pm$0.33 $\mu$m). The small value of standard deviation points to the independence of the track diameter on the difference between alpha particle energies originating from the $^{241} $Am source on one side and the good efficiency of the etching process on the other one.

\begin{figure}
 \centering
\includegraphics[width=\linewidth]{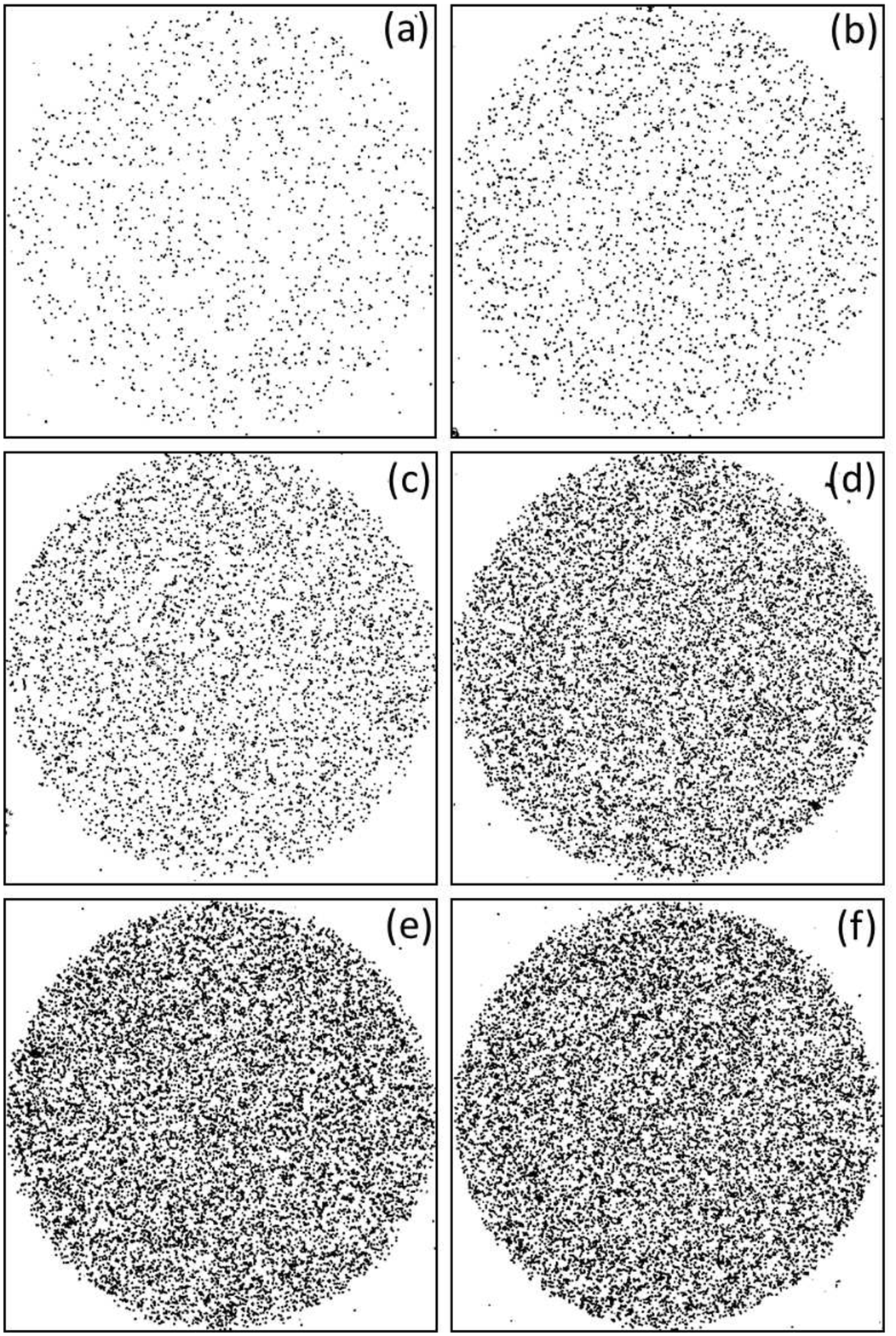}
 \caption{Photomicrographs of alpha particle tracks in PADC detector exposed for alpha particles of 4 MeV for different durations (a) 0.5 min, (b) 1 min., (c) 2 min., (d) 3 min., (e) 4 min, and (f) 5 min., PADC detector is chemically etched in 6.25 N NaOH at 70$\pm$1 ${}^{\circ}$C for 4 h. The area of the picture is 1583$\times$1583 $\mu$m${}^{2}$.}\label{Fig3}
\end{figure}

{The histogram in Fig}. \ref{Fig3} illustrates the variation of track density with time; these photomicrographs are for the samples chemically etched for 4 h. The area of each picture is 1583$\times$1583 $\mu$m${}^{2}$. The observed circular track diameters ranged from 8.4 to 9.1 $\mu$m. Several coalescent tracks are apparent in Figs. \ref{Fig3} (d-f) due to the expansion of the tracks to nearby ones. For low exposure time, the track densities are low, and no significant alpha tracks coalescence was observed; for instance, for irradiation time of 30 seconds, the alpha track density is (7.3$\pm$0.7)$\times10^{4}$ tracks.cm${}^{-2}$. For alpha particle irradiation time of 1 min, the alpha track density is (12.5$\pm$0.8)$\times10^{4}$ tracks cm${}^{-2}$; for a maximum exposure time of 5 minutes, the alpha track density amounts to (502$\pm$1)$\times10^{4}$ tracks cm${}^{-2}$.

A comparison of the response of the PADC detectors to exposed alpha particles for different durations and different etching times was undertaken by measurements of the track density in each of the samples listed in Table \ref{Tab1}. These samples of the PADC detectors are chemically etched for 2 h, 4 h, and 6 h in 6.25 N NaOH at 70$\pm$1 ${}^{\circ}$C, in which the response of the detector depends on the detectible track after etching.

\begin{figure}
 \centering
\includegraphics[width=\linewidth]{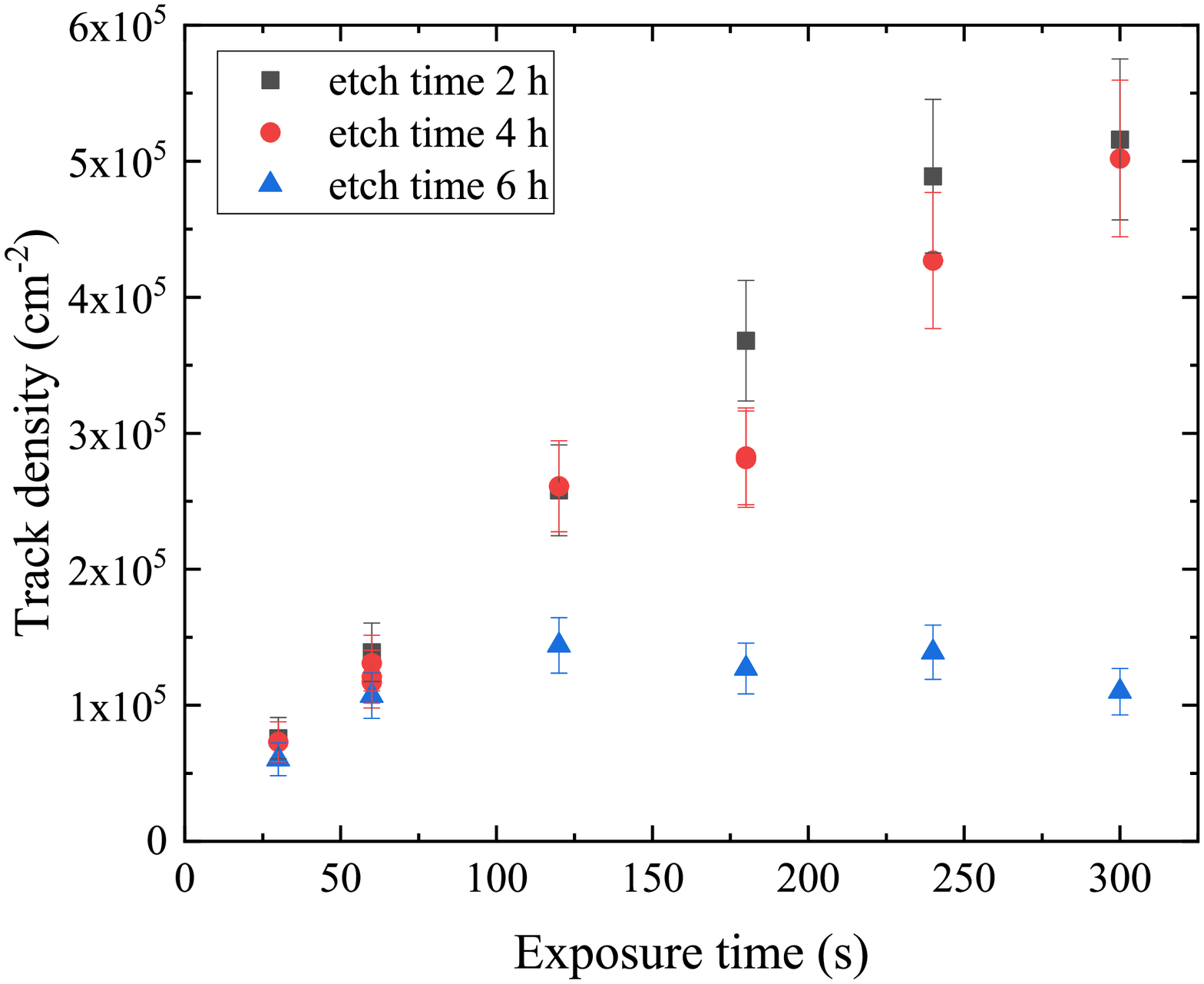}
 \caption{Alpha particle track density dependence on the exposure time at different etching times.}\label{fig4}
\end{figure}

As shown in Fig. \ref{fig4} for 4 h etching time, the tracks density and exposure time plot offered a non-linearity in registration efficiency. As exposure time increases, the diameters of alpha particle tracks are growing, and therefore coalescence and registered as one alpha reducing the track density, especially for longer exposure time. Similar non-linearity, nearly independent between the tracks density and exposure time, was evident for a larger etching time of 6 h.

For etching time of 2 h, the alpha particle track diameters were 4.9$\pm$0.2 $\mu$m, the alpha particle track densities were linearly correlated with irradiation time to alpha particle up to 240 s of exposure as depicted in Fig. \ref{fig4}. On the other hand, the maximum linearity for the samples etched for 4 h is 120 s of exposure time. In such circumstances, the linear registration of all spatially incident alpha particles grants minimum loss of information. Hence, patterns in samples f2, f4, and all samples etched for 6 h will give a biased conclusion on the extracted information.

\subsection{Randomness analysis}

Analyzing patterns is a well-established branch of computational science and information technology \cite{Howard2007275}. The most concerning abstraction is the point pattern analysis (PPA) \cite{Martinezsanchez2022106693} which involves the analysis of the spatial location of points in the multi-dimensional array (mostly two-dimensional). These analyses reveal deep laying information in such patterns. The divergence analysis is achieved by adopting the procedure of quadrat sampling to test its probability distributions and a statistical model to give predicted probabilities that may be compared with each of the individual probabilities in the observed frequency distribution. The theoretical probability distribution is obtained by making the sensible assumption of the randomness of the registration governing the evolution of the features in the pattern. From those assumptions, we deduced the probability distribution that will give the correct prediction of the frequency distribution of the quadrats. Finally, a comparison between the predicted probability distribution with the observed probability distribution obtained by sampling the pattern was made using Kullback-Leibler divergence based on the Shannon entropy hypothesis. There are no particular restrictions for the shape and size of a quadrat if the size is reasonable compared to the area under investigation. The selection of quadrat size is always an arbitrary procedure but may influence the subsequent interpretation of results. One of the most used treatments of quadrat size is the approach taken by Greig-Smith \cite{GreigSmith1952293}. On the basis that randomness at a variety of scales within a square quadrat census where the number of cells on each axis is some power of 2, based on the binary property of the Poisson distribution that it mean $\lambda $ equals its variance. However, in search for evidence of clustering at that scale, Greig-Smith has suggested that the size of quadrat at that scale will be related to the mean area of the pattern in which the test described here does not measure tendencies towards uniformity in the pattern. In the present work, we forced the quadrant area to follow the relation
\begin{equation} \label{Eq1}
A_{QS} =\sqrt{2} \frac{A}{N_{t} }
\end{equation}
Where A is the studied area and $N_{t} $ the total number of features in the whole pattern.

According to Poisson distribution, the null hypothesis of alpha tracks in SSNTD is the equal probability to hit any location in the exposed area, which implies that the number of hits is proportional to the detector area \textit{A} according to Poisson probability distribution. However, if there were clustering and dispersion in the pattern registered, the distribution would be different.

Poisson probability distribution ($q=q_{i} =q(x_{i} )$) of the number of features that will occur in a quadrat is
\begin{equation} \label{Eq2}
q(x_{i} )=\frac{\lambda _{i}^{x} }{\Gamma (x_{i} +1)} e^{-\lambda } ,
\end{equation}
which gives the random probability a number of $x_{i} $ events occur while being hit, $\lambda $ is the intensity function describes both the mean expected value and the variance of the distribution given from the relation;
\begin{equation} \label{Eq3}
\lambda =\frac{N_{P} }{N_{PQ} }
\end{equation}
where $N_{P} $ is the total number of features in the registered pattern within the investigated clip and $N_{PQ} $ is the number of quadrats to which the study area is divided. This analysis was undertaken for the samples a2, b2, c3, d2, e2, f2, a4, b4/1, b4/2, b4/3, c4, d4/1, e4, and f4, as labeled in Table \ref{Tab1}.

\begin{figure}
 \centering
\includegraphics[width=\linewidth]{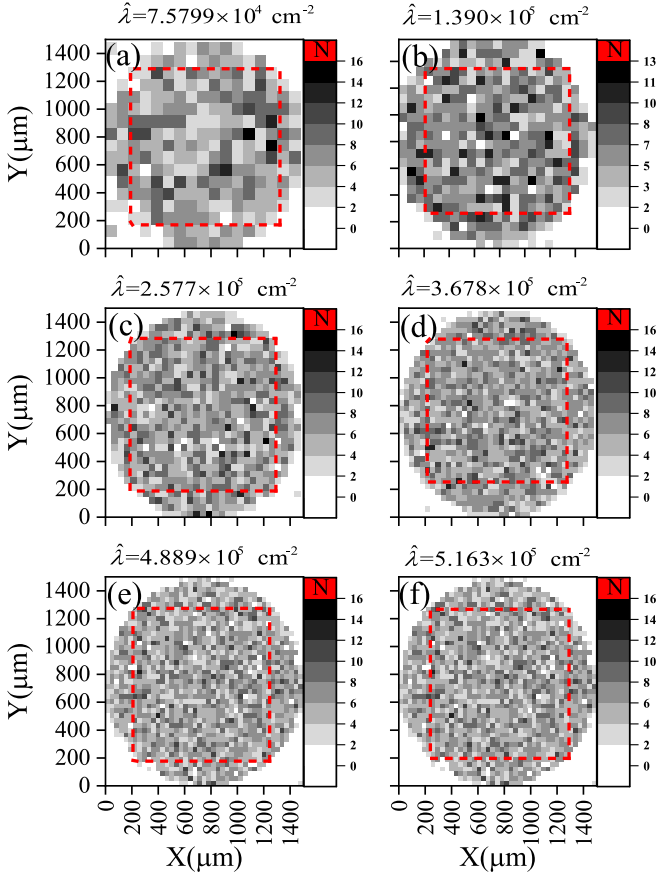}
 \caption{\textit{Heatmap} for the investigated local track densities of samples a2, b2, c3, d2, e2, and f2 after being etched for 2 h. The grayscale represents the number of features in each quadrate, $x_{i} $, while the number of quadrates increases progressively to fulfill Eq. \ref{Eq1}. The dashed squares (red color online) show clipped areas containing quadrats used for Poisson distribution analysis; see text.}\label{fig6}
\end{figure}

{The comparison in Fig}. \ref{fig6} shows a \textit{heatmap} for the number of features in each quadrate, $x_{i} =N$, counted in all investigated areas; The total number of quadrates depends on the condition in Eq. \ref{Eq1}. Detailed information is given in Table \ref{Tab2}. Generally, the statistics rely on the value of $\lambda $. So Eq. \ref{Eq1} grantee the closeness of the results upon comparison.

\begin{table}[p]
 \centering
 \caption{Detailed analysis of samples under study.}\label{Tab2}
\begin{tabular}{c|ccc|cccc} \hline \hline
 & \multicolumn{3}{c|}{Poisson Analysis} & \multicolumn{4}{c}{Entropy \& divergence} \\ \hline
Sample & $N_{P} $ & $N_{PQ} $ & $\lambda $ & $D_{KLD}^{(p||q)}$ & $\sigma _{X} $ & $\sigma _{Y} $ & $Q_{D} $ \\ \hline
a2 & 677 & 121 & 5.595 & 0.102 & 10.69 & 10.20 & 0.2470 \\
b2 & 1662 & 289 & 5.751 & 0.034 & 7.524 & 7.563 & 0.0195 \\
c2 & 2565 & 441 & 5.816 & 0.012 & 5.519 & 5.642 & 0.0615 \\
d2 & 3677 & 625 & 5.883 & 0.043 & 4.647 & 4.657 & 0.0050 \\
e2 & 5651 & 961 & 5.880 & 0.053 & 4.036 & 3.992 & 0.0220 \\
f2 & 5327 & 900 & 5.919 & 0.041 & 3.905 & 3.904 & 0.0005 \\
a4 & 667 & 121 & 5.512 & 0.118 & 10.38 & 10.85 & 0.2365 \\
b4/1 & 1396 & 225 & 6.084 & 0.051 & 7.988 & 8.242 & 0.1270 \\
b4/2 & 1486 & 256 & 5.804 & 0.064 & 8.337 & 8.317 & 0.0100 \\
b4/3 & 1274 & 225 & 5.662 & 0.050 & 7.859 & 7.885 & 0.0130 \\
c4 & 2589 & 441 & 5.871 & 0.013 & 5.487 & 5.615 & 0.0640 \\
d4/1 & 3004 & 529 & 5.678 & 0.029 & 5.321 & 5.271 & 0.0250 \\
e4 & 4717 & 784 & 6.017 & 0.056 & 4.302 & 4.255 & 0.0235 \\
f4 & 5158 & 900 & 5.725 & 0.120 & 3.927 & 3.991 & 0.0320 \\
\hline\hline
\end{tabular}
\end{table}

\begin{figure}
 \centering
\includegraphics[width=\linewidth]{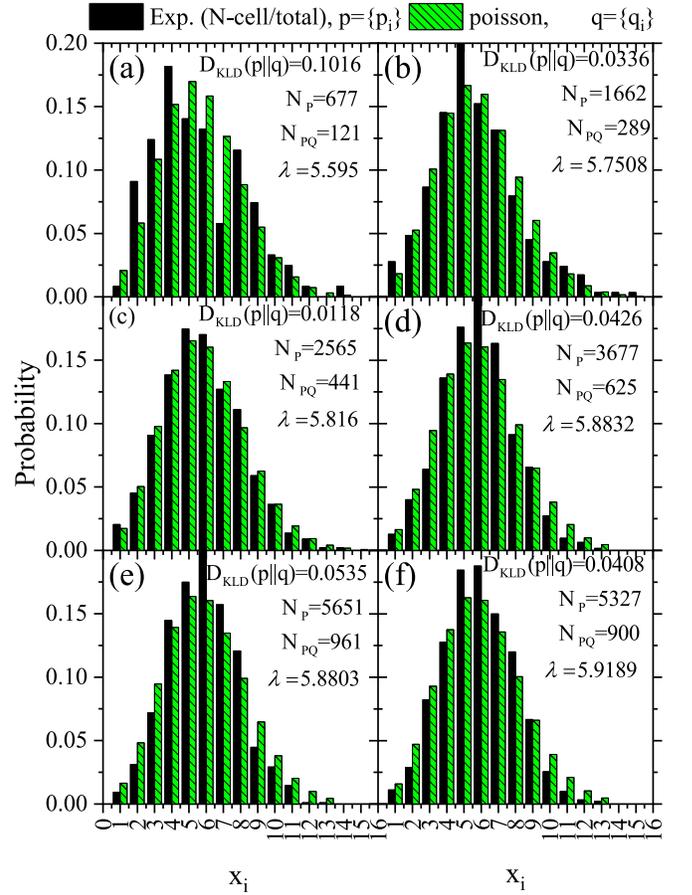}
 \caption{Probability histograms for the test hypothesis $p$ for samples a2, b2, c3, d2, e2, and f2 versus the null hypothesis $q$ versus the number of features $x_{i} $ within each quadrate. The null hypothesis is randomness based on Poisson distribution, in Eq. \ref{Eq2} assuming the same value of $\lambda $ as given in Table \ref{Tab2}.}\label{fig7}
\end{figure}

The histograms for the probability of a number of features $x_{i} $ within each quadrate deduced from the frequency statistics of the number of quadrats having a count $x_{i} $ divided by the total number of counts$N_{P} $ are shown in Fig. \ref{fig7}. For comparison, the Poisson distribution given in Eq. \ref{Eq2} was calculated assuming the exact value of the mean$\lambda $.

\begin{figure}
 \centering
\includegraphics[width=\linewidth]{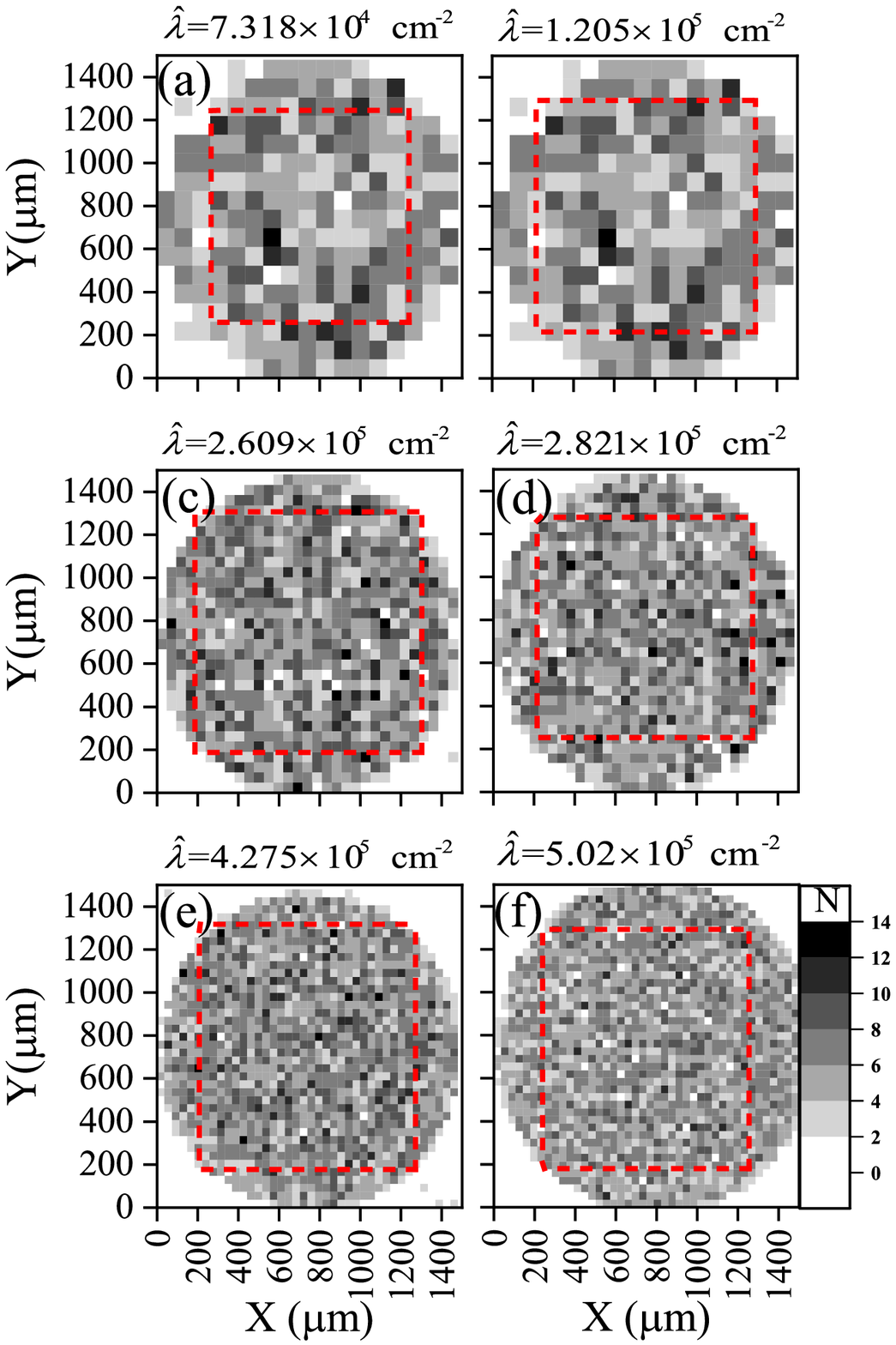}
 \caption{Constructed as in Fig. \ref{fig6} for the investigated local track densities of samples a4, b4/1, c4, d4/1, e4, and f4 after being etched for 4 h.}\label{fig8}
\end{figure}

\begin{figure}
 \centering
\includegraphics[width=\linewidth]{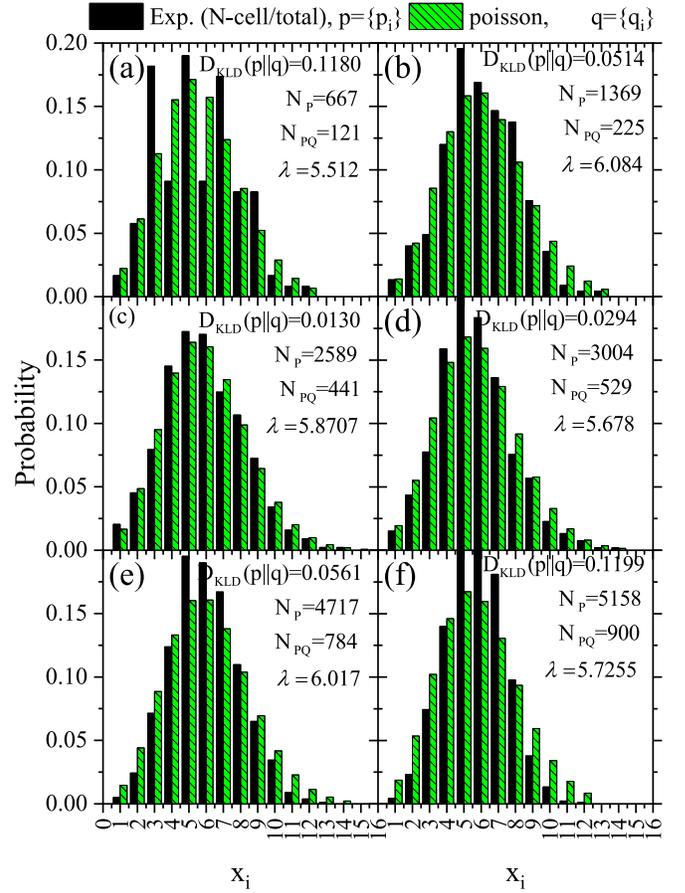}
 \caption{Probability histogram for the test hypothesis $p$ for samples a4, b4/1, c4, d4/1, e4, and f4 versus the null hypothesis $q$ versus the number of features $x_{i} $ within each quadrate. They were constructed as in Fig. \ref{fig7}.}\label{fig9}
\end{figure}

\begin{figure}
 \centering
 \includegraphics[width=\linewidth]{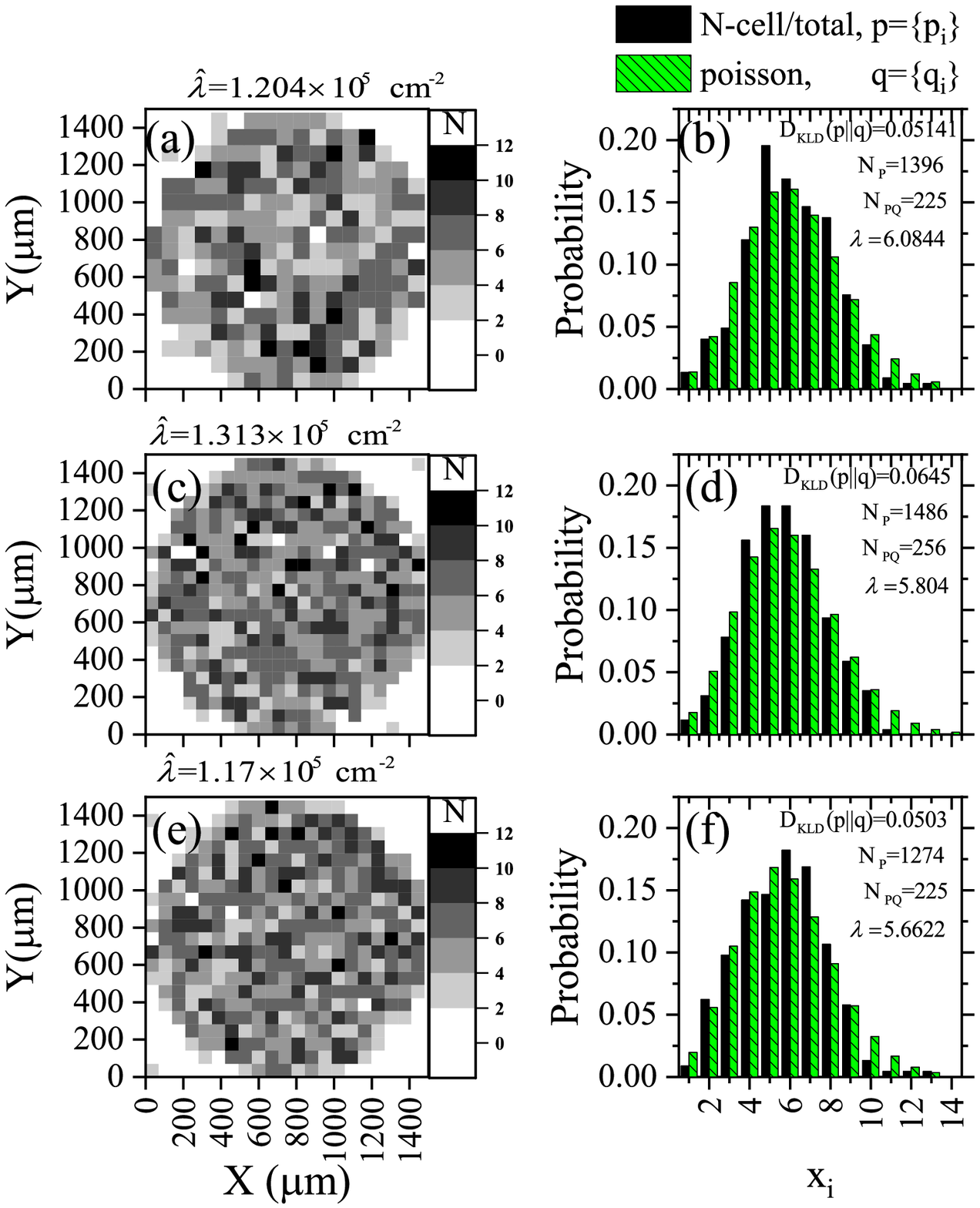}
 \caption{Comparison between \textit{heatmap} for the investigated local track densities of samples (a) b4/1, (c) b4/2, and (e) b4/3 and its associated probability histograms (b), (d), and (f), respectively. Constructed as figures \ref{fig7} and \ref{fig8}.}\label{fig10}
\end{figure}

The photomicrographs imaged for the samples a4, b4/1, b4/2, b4/3, c4, d4/1, e4, and f4 were analyzed using the same method. The results are illustrated in Figs. \ref{fig8}-\ref{fig10}.

\subsection{Entropy and divergence}

The amount of information concerning the variability of a random variable (uncertainty in randomness) of a statistical system of events directly indicates the system's Shannon entropy \cite{galar2017chapter}.
\begin{equation} \label{Eq4}
H(p)=-\frac{1}{\log N_{pQ} } \sum _{i}p_{i} \log p_{i}
\end{equation}
where $p=\{ p_{i} \} $ is the probability of the event in the $i\in \chi $ quadrat that belongs to the same probability space,$\chi $, of the observables. The most crucial concept in Shannon's Entropy is its ability to measure the extent to which the data are spread out over its possible values; lower entropy values refer to high information content and are most likely to develop a strong rule or correlation. For random data, the Shannon entropy value is equal to 1. In other terms, increased observability must lead to decreased uncertainty and entropy.

The difference between true random process and signal and more deterministic processes can be obtained using the Kullback-Leibler divergence (KLD) \cite{KullbackLeibler1951} divergence theoretical models \cite{Amati2002357}. KLD is a measure of dissimilarity between two probability distributions $p$ and $q=\{ q_{i} \} $ usually represents the probability distribution of data, the observations, and the probability distribution of its representing random model optimized for $p$. For the discrete case of data \cite{Cover1991Entropy}, KLD comprises
\begin{equation} \label{Eq5}
D_{{\rm KLD}} (p\parallel q)=\sum _{i}p_{i} \log \left(\frac{p_{i} }{q_{i} } \right)
\end{equation}
The positive value of $D_{{\rm KLD}} (p{\rm ||}q)$ represents the information gain achieved from \textit{p} instead of the random model \textit{q}. Based on Bayesian inference, $D_{{\rm KLD}} (p\parallel q)$ is the information gained upon measurement having posterior probability distribution $p$ compared to the priori known probability distribution $q$ or vice versa, the lost inference when forced random distribution $q$ is used instead of measured $p$ \cite{Burnham2004book}. The value of $D_{{\rm KLD}} (p\parallel q)$ goes to zero as the two probability distributions become the same.

The $D_{{\rm KLD}} (p\parallel q)$results are given in Table \ref{Tab2} and embedded within Figs. \ref{fig7}, \ref{fig9}, and \ref{fig10}. The last-mentioned figures are calculated based on the prior probabilities of random events based on Poisson distribution. The greater the prior uncertainty of such an occurrence, the greater the information gained if such a non-random event occurs. Criteria for defining an information statistic suggest that the measure would vary from zero to infinity and that the measure would be additive between independent events. \textit{The result showed embedded information within the track pattern}. Information could be extracted from patterns in samples a2, b2, c2, d2, e2, a4, b4/1, c4, d4/1, and e4. However, this information could be misleading due to other effects, as discussed below.

\subsection{Clustering and dispersion analysis }

Dispersion, skewness, and other major parameters can be clued from the central tendency analysis. While Clustering requires density analysis (including entropy and convergence) and distance analysis using pair correlation function (radial distribution function) and Ripley's K function \cite{Shakiba2022104531} involved in the spatial analysis method. The basic descriptive centric technique for a real data analysis is the featured center ($X_{c} $,$Y_{c} $) in which
\begin{equation} \label{Eq6}
X_{c} =\frac{1}{N\sum _{i=1}^{N}w_{i} } \sum _{i=1}^{N}w_{i} X_{i} ,
\end{equation}
\begin{equation} \label{Eq7}
Y_{c} =\frac{1}{N\sum _{i=1}^{N}w_{i} } \sum _{i=1}^{N}w_{i} Y_{i}
\end{equation}
$w_{i} $ is the weighting factor for the feature, which may be considered a reciprocal uncertainty of the existence of that point ($X_{i} $,$Y_{i} $) within the center area of the feature. For definitely shaped points, $w_{i} $=1, the variance of the distribution of the data may be different in the directions X and Y,
\begin{equation} \label{Eq8}
\sigma _{Y}^{2} =\frac{1}{(N-1)\sum _{i=1}^{N}w_{i} } \sum _{i=1}^{N}w_{i} \left(Y_{i} -Y_{c} \right)^{2}
\end{equation}
\begin{equation} \label{Eq9}
\sigma _{X}^{2} =\frac{1}{(N-1)\sum _{i=1}^{N}w_{i} } \sum _{i=1}^{N}w_{i} \left(X_{i} -X_{c} \right)^{2} .
\end{equation}
The distribution deviation is determined by the relation
\begin{equation} \label{Eq10}
\sigma _{D}^{2} =\frac{\sigma _{X}^{2} +\sigma _{Y}^{2} }{2} ,
\end{equation}
while the quality of the distribution is determined by the relation
\begin{equation} \label{Eq11}
Q_{D}^{} =\left|\frac{\sigma _{X}^{2} -\sigma _{Y}^{2} }{2} \right|.
\end{equation}
The standard deviation in two dimensions is defined by
\begin{equation} \label{Eq12}
\sigma ^{2} =\frac{\sum _{i=1}^{N}w_{i} \left(\left(X_{i} -X_{c} \right)^{2} +\left(Y_{i} -Y_{c} \right)^{2} \right)}{(N-2)\sum _{i=1}^{N}w_{i} }  .
\end{equation}
The (\textit{N}-2) provides an unbiased estimate of standard distance since there are two constants related to a real deviation. Note that $\sigma ^{2} \ne \sigma _{D}^{2} $ if a circular clip of pattern was taken whether or not $Q_{D} \ne 0$.

As shown in Table \ref{Tab2}, the photomicrograph patterns of the alpha particle tracks do not offer a uniform spatial distribution around the center of the data. The $Q_{D} $values span a range from 0.005 to 0.247. A large value of $Q_{D} $ at low exposure time is attributed to a limited number of registered tracks in the detectors to attain random data. \textit{Hence, patterns a2, b2, a4, and b4/1 contain remnant information of randomness despite a2, and a4 have large values of entropy divergence.}\textbf{ } Similarly, at a large exposure time of 300 s, tracks coalescence may disturb the gained information. At the intermediate track densities, the value of $Q_{D} $ begins to reach 0, the nominal value of random track registration. Conversely, its value may increase due to the accumulation of clustering information within the alpha particle tracks.

\subsection{The empirical K-function}

The empirical distribution function is the pairwise distances used to search for anomalies in the feature patterns. The second moment of this distribution function is the differential Radial Distribution Function (RDF) as a function of distance r. Our focus is on the distance or spacing between features in the registered pattern. Each ordered pair of points had a measured distance $d_{i,j} =||r_{i} -r_{j} ||$ which may contain the information about the alpha particles' spatial pattern.

\begin{figure}
 \centering
\includegraphics[width=\linewidth]{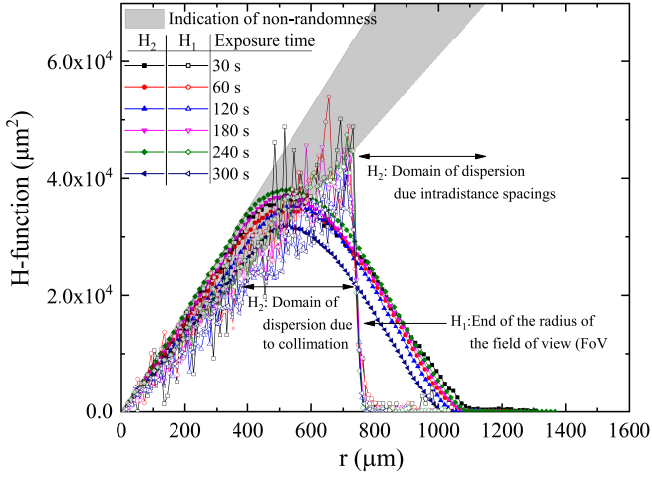}
 \caption{Plot of the Radial Distribution Function $H_{1} (r_{i} )$and $H_{2} (r_{i} )$for samples a2, b2, c2, d2, e2, and f2 after being etched for 2 h.}\label{fig11}
\end{figure}

\begin{figure}
 \centering
\includegraphics[width=\linewidth]{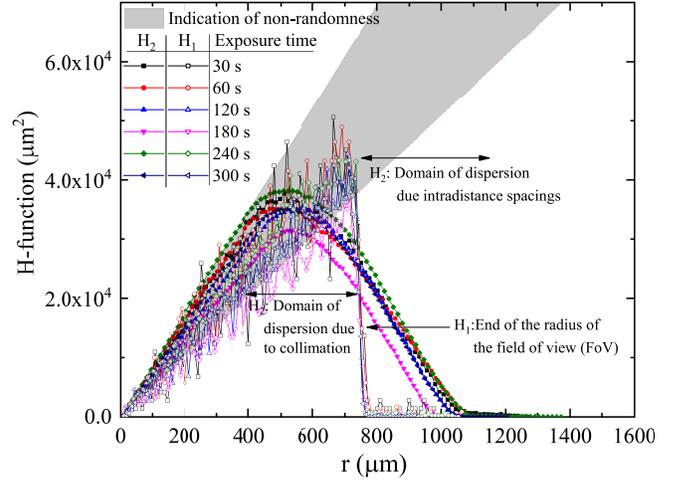}
 \caption{Plot of the Radial Distribution Function $H_{1} (r_{i} )$and $H_{2} (r_{i} )$for samples a4, b4/1, c4, d4/1, e4, and f4 after being etched for 4 h.}\label{fig12}
\end{figure}

There are two different definitions used in the present work of the RDF as a function of distance r, first,
\begin{equation} \label{Eq13}
H_{1} (r_{i} )=\frac{1}{\hat{\lambda }} \sum _{i=1} \vec{1}(r_{i-1} <d_{i,c} <r_{i} )
\end{equation}
$d_{i,c} $ is the distance between the features labeled $i$ and the center of the data, $\vec{1}$(condition) is the indicator function for the satisfaction of the condition and $\hat{\lambda }$ is the average number of features per unit area. In this case, the cutoff radial distance is the radius of the clipped pattern (denoted $d_{c} )$. Also, as a function of pair distances
\begin{equation} \label{Eq14}
H_{2} (r_{i} )= N\cfrac{\sum _{i}\sum _{j\ne i} \vec{1}(d_{i,j} <r_{i}) }{\hat{\lambda}({\rm No.\; of\; interdistances})}
\end{equation}
The condition $d_{i,j} <d_{c} -r_{i} $ was introduced to enforce the calculation to run only to the pattern within the clipped circle and reduce the edge effect of the data counting. Value is normalized to the new counted features. The plot of the functions $H_{1} (r_{i} )$and $H_{2} (r_{i} )$ is shown in Fig. \ref{fig11} and \ref{fig12}.

The variation of the central distribution function reveals a sort of correlation in the pattern, which may be symmetric around the center of the {registered} pattern. {Such central tendency is a consequence of the existence of the collimator and the inverse-square law in which the emitted alpha particles from the source may be inclined along the diagonal of the collimator rather than parallel to its axis}.  The pair RDF cannot detect such a pattern due to its moving average nature.

Here, the radial distribution function $H_{1} (r_{i} )$and $H_{2} (r_{i} )$gave two crucial pieces of information. $H_{1} (r_{i} )$gives the distribution around the center of the data from equations 6 and 7, in which the central symmetry of the function compensates for the effect of the non-randomness of the data. The data is truncated at the end of the field-of-view (FoV). $H_{2} (r_{i} )$, on the other hand, is a pair distribution function that is sensitive to clustering and dispersion of the pattern and the edge effect. Collimation of alpha particles on a determined region on the detector causes two effects: (1) the pairs near the edge of the area have fewer neighbors from one side, which lesser the value of $H_{2} (r_{i} )$near the end of the FoV. (2) remote intradistant neighbors from the other directions gave a value of $H_{2} (r_{i} )$at distances greater than the end of FoV. The difference between these orders, within a radius of about 1/3 of the diameter of the data, is another clue to the existence of clustering or dispersion in the pattern, as shown in Fig. \ref{fig11}. Similar behavior was observed for these samples etched for 4 h (see Fig. \ref{fig12}.)

\begin{figure}
 \centering
 \includegraphics[width=\linewidth]{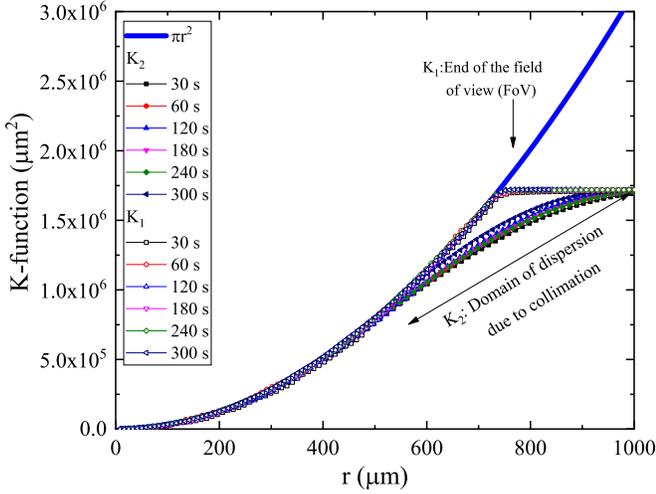}
 \caption{Plot of the cumulative Radial Distribution Function $K_{1} (r)$and $K_{2} (r)$for samples a2, b2, c2, d2, e2, and f2. The values of $\pi r^{2} $ were added for comparison.}\label{fig13}
\end{figure}

\begin{figure}
 \centering
 \includegraphics[width=\linewidth]{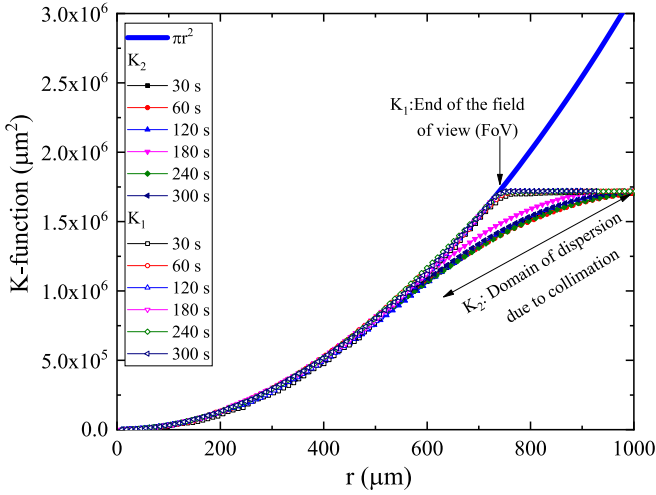}
 \caption{Plot of the cumulative Radial Distribution Function $K_{1} (r)$and $K_{2} (r)$for samples a4, b4/1, c4, d4/1, e4, and f4. The values of $\pi r^{2} $ were added for comparison.}\label{fig14}
\end{figure}

The K-function, on the other hand, is a Cumulative Radial Distribution Function (CRDF), obtained based on $H_{1} (r_{i} )$ and $H_{2} (r_{i} )$, as
\begin{equation} \label{Eq15}
K_{1} (r)=\sum _{\begin{array}{c} {i=1} \\ {r_{i} \le r} \end{array}} w_{i,c} H_{1} (r_{i} )
\end{equation}
\begin{equation} \label{Eq16}
K_{2} (r)=\sum _{\begin{array}{c} {i=1} \\ {r_{i} \le r} \end{array}} w_{i,c} H_{2} (r_{i} )
\end{equation}
$w_{i,c} $ is a generally used edge-corrected estimator. The weight function has a value between 0 and 1 that provides higher weight to the points needing the center of the investigated area rather than the points at the edge. In the present work, we shall use $w_{i,c} =1$. i.e.
\begin{equation} \label{Eq17}
K_{1} (r)=\frac{1}{\hat{\lambda }} \sum _{i} 1(d_{i,c} <r)
\end{equation}
\begin{equation} \label{Eq18}
K_{2} (r)=N\cfrac{ \sum _{i}\sum _{j\ne i} 1(d_{i,j} <r_{i} ;d_{i,j} <d_{c} -r_{i} )}{\hat{\lambda }({\rm No.\; of\; points})}
\end{equation}
The plot of the functions $K_{1} (r_{i} )$and $K_{2} (r_{i} )$ are shown in Figs. \ref{fig13} and \ref{fig14}. It is obvious that the $K_{1{\rm \; or\; }2} $-functions and $H_{2} $ -function do not uniquely define the pattern. Still, they can be used to detect if there were a \textit{direct interaction} between processes causing the pattern, i.e\textit{.}, two different patterns may have the same K-function, see Refs. \cite{Lotwick1984575,Baddeley2021100435}.

In this aspect, The null hypothesis of the K-function is that the number of features lying closer than a distance r has expected value $K_{1{\rm \; or\; }2} (r)$, i.e., the variation as $\pi r^{2} $ and deviations from this expectation indicate scales of clustering and/or dispersion \cite{Dixon2002inbook,Dixon2014inbook}. An inhibited process causes a lake of formation of the feature and will usually have $K_{1{\rm \; or\; }2} (r)<\pi r^{2} $, while an enhanced process causes clustered feature and will have $K_{2} (r)>\pi r^{2} $, for appropriate values of r. While $K_{1} $ is related to the nearest-center distribution and related mainly to the anisotropy of the radial signature of the features, $K_{2} $ is associated with the nearest-neighbor distribution and is related to the none-stationary processes causing the feature, also known as Ripley's K \cite{Ripley1976255,Ripley1977172}.

Consequently, trends of \textbf{$K_{1} (r)$}follow the $\pi r^{2} $ trend to the end of FoV while the trends of \textbf{$K_{2} (r)$}follow the $\pi r^{2} $ trend up to 1/3 of the diameter of the collimator (2/3 of the radius to the end of FoV).

\begin{figure}
 \centering
 \includegraphics[width=\linewidth]{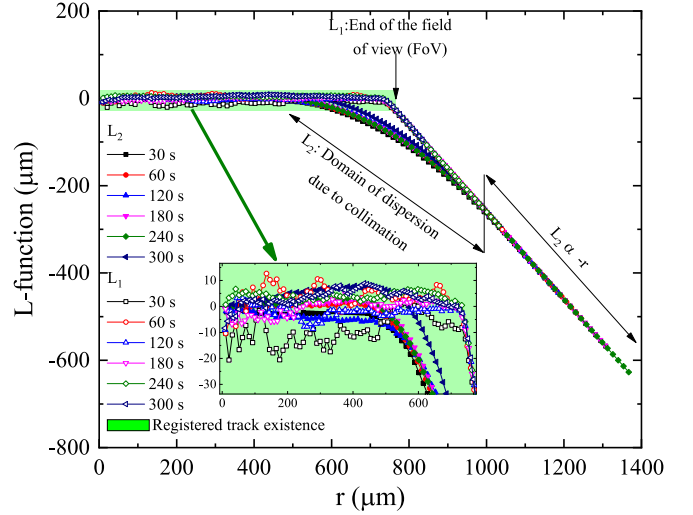}
 \caption{Plot of the L-Functions $L_{1} (r)$and $L_{2} (r)$for samples a2, b2, c2, d2, e2, and f2.}\label{fig15}
\end{figure}

\begin{figure}
 \centering
\includegraphics[width=\linewidth]{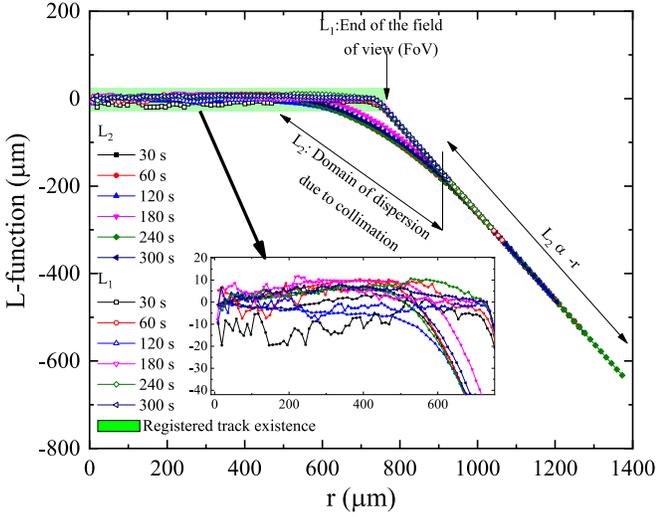}
 \caption{Plot of the L-Functions $L_{1} (r)$and $L_{2} (r)$for samples a4, b4/1, c4, d4/1, e4, and f4.}\label{fig16}
\end{figure}

Because of the difficulty of comparison, we consider the difference L-Functions
\begin{equation} \label{Eq19}
L_{1} (r)=\sqrt{\frac{k_{1} (r)}{\pi } } -r
\end{equation}
\begin{equation} \label{Eq20}
L_{2} (r)=\sqrt{\frac{k_{2} (r)}{\pi } } -r
\end{equation}
These functions are plotted in Figs. \ref{fig15} and \ref{fig16}. Since the L-function has the dimension of distance, the confidence interval is just the confidence interval of each feature, i.e., the 0.3 $\mu$m deduced from the analysis of track diameter. The values of L-function greater than this value represent correlation among clusters, which occur at about around 200-500 $\mu$m.

From these results, the essential \textit{reasonable information could be extracted from samples c2 and c4.} However, c4 has more features, as depicted in Table \ref{Tab1}. Hence, sample c4 was the most candidate pattern to extract information.

\subsection{Proximity Analysis }

\begin{figure}
 \centering
\includegraphics[width=\linewidth]{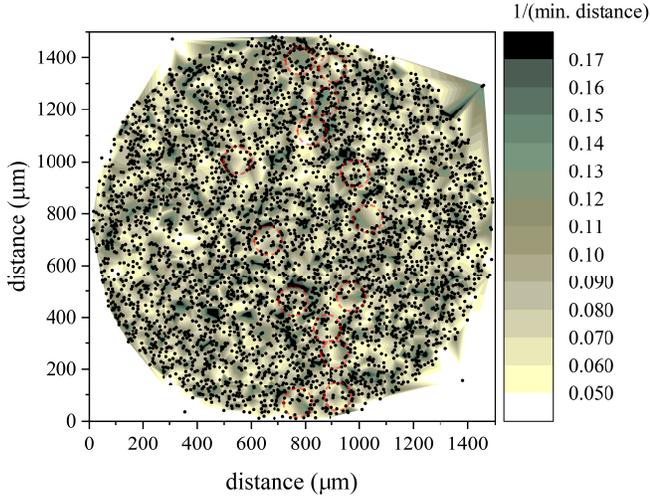}
 \caption{Proximity analysis diagram based on a buffer of reciprocal NND. Semitransparent circles show some correlated features.
}\label{fig17}
\end{figure}

To find out what is near or within a certain distance of one or more features, we use a common geographic information system process that includes a buffer as a tool that creates a new feature class of buffer polygons around a specified input feature based on some factor. Our factor is the reciprocal of nearest neighbor distance (NND). The value of NND is inversely proportional to the density of the features and tells much about whether data points are clustered or dispersed. {In this aspect, Fig}. \ref{fig17} shows the proximity analysis based on that buffer obtained from the analysis of radial distribution.

The data of the proximity analysis in Fig. \ref{fig17} showed long-range correlational symmetry between regions of clusters. Semitransparent circles show some correlated features having a dimension of diameter 100 micrometers. {Similar clustering in the registered tracks has been observed in all samples. These are obvious in the heatmaps in Figs. \ref{fig6}, \ref{fig8}, and \ref{fig10}}. The origin of such a correlation is unknown. The present results throw doubt on the validity on the random model of estimating the trajectory of alpha particles, especially in the case of use of isotopic source as an initiator for the neutron emitting reactions (e.g. $^{241}$AmBe neutron source \cite{TohamyElmaghrabyComsan2019162387}). In radiation detection, the phenomenology of assessing the radioactivity may be influenced by the configuration mixing between particle decay and its interaction \cite{Elmaghraby2017PiP} and interfere with the possible time varying-decay rates \cite{Jenkins201350,Sturrock201442,Sturrock20168,Aharmim2007045502,MilianSanchez20208525,Cooper2009267,Bellotti2013116,Alexeyev201323,Pomme2016281,Bikit201338,Pomme20201093}.

\section{Conclusion}

The phenomenology of charged particle emission upon the nuclear decay showed a spatial correlation among trajectories of the particle. {The statistical inference methods, specifically the information entropy, track proximity analysis, and Kullback-Leibler divergence, was used to track the information gained/lost due to the alpha particles trajectories. The alpha particle from an ${}^{241}$Am source have to pass through a thin film of gold and registered at the end of its trajectory on the SSNTD}. All sources of a possible influence on the results were tested and eliminated from the experiment and the analysis. This includes the authentication of linearity in registration efficiency with exposure time to avoid coalescence of registered tracks, gravitational alinement, electromagnetic interferences, etc; so that all experimental parameters were optimized to identify the best conditions of exposure and etching to avoid misleading results. The adaptive quadrates analysis of the spatial data showed that the trajectories do not follow the null hypothesis of Poisson trajectories upon hitting the detector surface. Entropy and divergence analysis of the quadrate data with the null hypothesis of Poisson distribution.  Entropy analysis showed information gained upon registration of alpha particles tracks on SSNTD.  The clustering and dispersion analyses were undertaken with central deviation tendency, empirical K-function, radial distribution analysis, and proximity analysis. Results showed the existence of pattern information within the registered tracks that may be attributed to the structure of the source materials or coherence among emitted alpha particles. The source of correlation may be any process from the creation of the alpha particle inside the ${}^{241}$Am nucleus, the electron density in the source material {(the deposited salt of Am) or its crystal structure, the electromagnetic interactions with atoms in encapsulation gold film, or} even resonance with an ambient electromagnetic field. {One of the main consequences from the presented results is the possibility of influencing the accuracy of the particle detectors and other experimental techniques that are used in high-energy physics and nuclear physics, in which the calibration is based on the randomness hypothesis}.


%

\end{document}